\documentclass[11pt]{article}
\usepackage{aaspp4}
\slugcomment{draft of \today}

\begin{document}

\font\twelvei = cmmi10 scaled\magstep1 
       \font\teni = cmmi10 \font\seveni = cmmi7
\font\mbf = cmmib10 scaled\magstep1
       \font\mbfs = cmmib10 \font\mbfss = cmmib10 scaled 833
\font\msybf = cmbsy10 scaled\magstep1
       \font\msybfs = cmbsy10 \font\msybfss = cmbsy10 scaled 833
\textfont1 = \twelvei
       \scriptfont1 = \twelvei \scriptscriptfont1 = \teni
       \def\mit{\fam1 }
\textfont9 = \mbf
       \scriptfont9 = \mbfs \scriptscriptfont9 = \mbfss
       \def\bmit{\fam9 }
\textfont10 = \msybf
       \scriptfont10 = \msybfs \scriptscriptfont10 = \msybfss
       \def\bmsy{\fam10 }

\def\etal{{\it et al.~}}
\def\cf{{\it cf.~}}
\def\eg{{\it e.g.,~}}
\def\ie{{\it i.e.,~}}
\def\lsim{\raise0.3ex\hbox{$<$}\kern-0.75em{\lower0.65ex\hbox{$\sim$}}} 
\def\gsim{\raise0.3ex\hbox{$>$}\kern-0.75em{\lower0.65ex\hbox{$\sim$}}} 
\def\kms{~{\rm km~s^{-1}}}
\def\kmsmpc{~{$\rm km~s^{-1}~Mpc^{-1}$}}
\def\cm3{~{\rm cm^{-3}}}
\def\yr{~{\rm yr}}
\def\Msun{~{\rm M}_{\sun}}

% My def.
\def\hinv{$h^{-1}$}
\def\lcdm{$\Lambda$CDM~}
\def\va{$v_s$~}
\def\ra{$R_s$~}
\def\tx{$T_x$~}
\def\tcs{\tau_{cs}}
\def\ssp{c_{si}}
\def\ltsima{$\; \buildrel < \over \sim \;$}
\def\simlt{\lower.5ex\hbox{\ltsima}}
\def\gtsima{$\; \buildrel > \over \sim \;$}
\def\simgt{\lower.5ex\hbox{\gtsima}}

\title{Properties of Cosmic Shock Waves in Large Scale Structure
Formation\altaffilmark{5}}

\author{  Francesco Miniati     \altaffilmark{1},
          Dongsu Ryu            \altaffilmark{2}, 
          Hyesung Kang          \altaffilmark{3},
          T. W. Jones           \altaffilmark{1},
          Renyue Cen             \altaffilmark{4} \\
      and Jeremiah P. Ostriker  \altaffilmark{4}}

\altaffiltext{1}{School of Physics and Astronomy, University of Minnesota,
    Minneapolis, MN 55455}
\altaffiltext{2}{Department of Astronomy \& Space Science, Chungnam National
    University, Daejeon, 305-764 Korea}
\altaffiltext{3}{Department of Earth Science, Pusan National University,
    Pusan, 609-735 Korea}
\altaffiltext{4}{Princeton University Observatory, Princeton, NJ 08544}
\altaffiltext{5}{To appear in {\it The Astrophysical Journal}}

\begin{abstract}

We have examined the properties of shock waves in
simulations of large scale structure formation. Two cosmological scenarios
have been considered: a standard cold dark matter model with $\Omega_M=1$
(SCDM) and a cold dark matter model with cosmological constant and
$\Omega_M+\Omega_{\Lambda}=1$ ($\Lambda$CDM) having $\Omega_\Lambda=0.55$. 
Large-scale shocks result from accretion onto sheets, filaments and
knots of mass distribution on a scale of order of $\sim 5$\hinv Mpc 
in both scenarios.
Energetic motions, partly residual of past 
accretion processes and partly caused by current
asymmetric inflow along filaments, end up
generating additional shocks. These extend on a scale of order of
$\sim1$\hinv Mpc and envelop and penetrate deep inside the 
clusters.
Also collisions between substructures inside clusters form merger shocks. 
Consequently, the topology of the shocks is very complex and highly
connected. During cosmic evolution the comoving shock surface density
decreases, reflecting the ongoing structure merger process
in both scenarios.

Accretion shocks have very high Mach numbers, typically between
10 and a few $\times 10^3$, when photo-heating of
the pre-shock gas is not included. The characteristic shock 
velocity is of order $v_{sh}(z)=H(z)\lambda_{NL}(z)$, where 
$\lambda_{NL}(z)$ is the wavelength scale of the nonlinear perturbation
at the given epoch.
However, the Mach number for
merger and flow shocks (which occur within clusters) is usually smaller,
in the range $\sim 3 - 10$, corresponding to 
the fact that the intracluster gas is hot (i.e., already shock heated). 
Statistical fits of shock velocities around clusters as a
function of cluster temperature give power-law functions in accord with
those predicted by one-dimensional solutions. On the other hand, a very
different result is obtained for the shock radius, reflecting
extremely complex shock structures surrounding clusters of galaxies in 
three-dimensional simulations.
The amount of in-flowing kinetic 
energy across the shocks around clusters, which represents
the power available for
cosmic-ray acceleration, is comparable to the cluster X-ray 
luminosity emitted from a central region of radius 0.5 \hinv Mpc.
Considering their large size and long lifetimes, those shocks are
potentially interesting sites for cosmic-ray acceleration, if modest
magnetic fields exist within them. 

\end{abstract}

\keywords{acceleration of particles --- cosmology: large-scale structure
of universe --- galaxies: clusters: general --- methods: numerical ---
shock waves --- X-rays: galaxies }

% \clearpage

\section{Introduction}

``Cosmic shock waves'', formed in the course of large-scale structure 
formation, can contribute important roles in cosmology. They include
external {\it accretion shocks} as well as {\it merger} and 
{\it flow shocks} internal to galaxy clusters.
The pristine cosmic plasma accreting onto the large scale structure is 
deflected from the Hubble flow and first processed by {\it accretion shocks}
(see \eg Ryu \& Kang 1997b). Evidence for their existence might be
inferred from the observation of hot gas in the intracluster medium (ICM).
In the commonly accepted paradigm for structure formation in the universe,
gas accreting onto cosmic filaments and clusters of galaxies (GCs) has a
typical bulk velocity up to $\sim {\rm a~few}~10^3 \kms$. This gas
is then shock-heated to temperatures ranging from $10^5-10^7$ K in filaments
and up to $10^7- 10^8$ K in GCs (\eg Kang \etal 1994a, KCOR94 hereafter;
Cen \& Ostriker 1994, CO94 hereafter; Cen \& Ostriker 1999a).
{\it Merger shocks} are produced during the mergers of sub-structures within a 
galaxy cluster and propagate through the hot ICM. In addition, 
during such a process, accretion shocks associated with the merging
units also propagate through the ICM. Together with merging shocks they form a
complex structure
that can survive for long times inside the ICM after the end of the merger, 
because of the continuous gas inflow through filaments and sheets. We refer to
these as {\it flow shocks} (see \S \ref{morp.s}).
There is now substantial observational evidence for temperature
structure in clusters due to internal shock waves; 
these are appear to be mostly produced by 
merger events (\eg Markevitch \etal 1998; Donnelly \etal 1999
and references therein) although recently some evidence might have appeared
for flow shocks, too (\cite{e98}, \S \ref{dc.s}).

Among other reasons for
interest in cosmic shocks is their ability to efficiently accelerate
particles to relativistic energies (\cf Blandford \& Ostriker 1978, 1980;
and Jones \& Ellison 1991 for a recent review of this subject). In fact,
relativistic cosmic-ray (CR) electrons have been observed in GCs
through their synchrotron emission (\eg Kim \etal 1989; 
Giovannini \etal 1993; Deiss \etal 1997). Extended sources of 
synchrotron radiation are commonly observed with spatial distribution similar 
to that of the thermal X-ray emission (see \eg Liang 1999). Although the 
cooling time for such CR electrons is much shorter than the cluster ages, 
explicit signatures of 
particle aging are rare in the spectra of the observed sources. Since 
individual cluster galaxies are unlikely to replenish the ICM adequately 
with populations of relativistic particles, an efficient mechanism 
for extended particle acceleration is probably required to understand 
the CR electron replenishment. 

Moreover, recently the EUVE satellite has revealed that many clusters possess 
an excess of extreme ultra-violet (EUV) radiation compared to what is 
expected from the hot, thermal X-ray emitting ICM (\eg Lieu \etal 1996; 
Fabian 1996; Mittaz \etal 1997; Kaastra 1998). 
Further evidence for nonthermal activity in the ICM comes 
from detection of radiation in excess to thermal emission in the hard 
X-ray band above $\sim 10$ KeV (\eg Henriksen 1998; Fusco-Femiano \etal 
1999; Valinia \etal 1999; Sarazin 1999). 
The mechanism proposed for the origin of these components is the 
inverse-Compton (IC) scattering of cosmic microwave background photons
by CR electrons, although it is not clear if the same electron population 
is responsible for producing both the EUV excess and the hard X-ray excess 
(Ensslin \etal 1999).
Such detections suggest the possibility that nonthermal activities in the 
ICM are much higher than previously expected (Sarazin \& Lieu 1998;
Lieu \etal 1999).

Cosmic shock waves should be capable of
accelerating CRs electrons responsible for the above emissions. 
However, at the same time, CR protons are produced.
It is possible then, although not established yet, that if the above
interpretation for the nonthermal radiation is correct, the CR
protons produced at these shocks and accumulated throughout the cosmological
evolution could provide a substantial fraction of the total 
pressure in GCs (Sarazin \& Lieu 1998; Lieu \etal 1999). 
It is clear that if the CR pressure 
was ever comparable to the thermal pressure during the evolution of the 
universe, that would have a profound impact on cosmology. For instance,
structure formation is heavily used as a probe for discriminating among 
cosmological models (\eg Carlberg \etal 1997; 
Bahcall \& Fan 1998) and hydrostatic equilibrium of the thermal ICM gas in
the potential well of the total cluster mass is commonly assumed in
order to derive GC masses (\eg White \etal 1993; Evrard 1997).
The presence of a nonthermal component obviously would alter the results in
proportion to its relative importance.
Furthermore some additional
source of pressure is clearly required in GCs over that produced
by adiabatic hydrodynamics both to produce the correct density 
profiles (\eg Evrard 1990; Navarro, Frenk \& White 1995) 
and to prevent catastrophic cooling flows (Suginohara
\& Ostriker 1999). Cosmic ray pressure in the inner parts of GCs
may thus play a vital role in the hydrodynamic equilibrium of
these systems.

Accretion shocks were also proposed as sites for acceleration of high 
energy CRs, protons and heavy nuclei, up to $10^{18} - 10^{19}$eV 
(Kang \etal 1996, 1997). In fact, given
the large velocity of the accretion flows and the large
size and long lifetimes of the associated shocks, such energies
would be achievable through 
``cross-field'' diffusion in perpendicular magnetohydrodynamic shocks.
Magnetic fields in the ICM of GCs have been
observed with strengths of the order of a few $\times 0.1\mu$G
(\eg Kim \etal 1989; Fusco-Femiano \etal 1999; Molendi \etal 1999).
Outside GCs constraints on rotational measure 
from quasars impose an upper limit of $\sim$nG, based on the 
assumption of regularly alternating magnetic field (see \eg Kronberg 1994). 
However, this limit can be shifted to higher values 
if a realistic distribution of magnetic field associated with cosmic 
structures is assumed. On such basis, Ryu \etal (1998) and Blasi \etal (1999) 
claimed a new {\it upper limit}
\ltsima 1$\mu$G, at least along cosmic structures.

Cosmic shock waves could serve also as sites for the generation of weak 
seeds of magnetic field by the Biermann battery mechanism.
It was proposed that these seeds could be amplified to strong magnetic 
field of up to $\sim\mu{\rm G}$ in clusters if flows there can be described as
Kolmogoroff turbulence (Kulsrud \etal 1997). However, further development
into coherent magnetic field is unclear, since there is as yet no detailed
theory capable of describing this process (see, \eg Chandran 1997).

Additional roles which shock waves may play in cosmology have been 
explored by a number of authors (\eg Ryu \& Kang 1997a, RK97 hereafter; 
Quilis \etal 1998). RK97 compared analytical self-similar solutions for 
cluster formation in  the Einstein-de Sitter universe
(Bertschinger 1985) with one-dimensional numerical 
simulations in low density universes with/without a cosmological constant
($0.1 < \Omega_M < 1$), where the properties of the accretion flow are 
related with the cluster's mass, radius and temperature. 
The major conclusion was a possibly testable prediction about the 
difference in the accretion flow in different cosmological models. 
In particular, the accretion velocity onto clusters of a given mass or 
radius in low density universes is smaller by up to 45 \% and 65 \% 
respectively compared to that in the Einstein-de Sitter universe.

In the present paper we focus on the quantitative properties of large scale 
shocks produced by gas during the formation of cosmic structures. For this,
the simulation data described in KCOR94 and CO94 have been used. The roles
played by those shocks, especially with regard to 
CR acceleration and magnetic field
generation and their consequences on cosmology, will be studied in
future work. Details of our data analysis are described in \S \ref{num2.s}. 
In particular, we computed the velocity, Mach number, radius of 
shocks and kinetic energy flow across them, which were not 
studied in previous works, and calculated their correlation with the 
cluster core temperature. The results are reported in \S \ref{res.s}. 
Finally, \S \ref{dc.s} concludes with a discussion.

\section{Numerics} \label{num.s}

\subsection{Simulations \& Data} \label{num1.s} 

We study the properties of shock waves associated with 
the large scale structure of the universe in numerical simulations.
In particular we investigate
the velocity and radius of shocks in terms of cluster's X-ray 
temperature. The latter, in fact, is the 
most reliably reproducible quantity in numerical 
simulations (\eg Kang \etal 1994b; Frenk \etal 1999) and has been recently
measured with satellite observations (\eg Arnaud 1994; Mushotzky \& 
Scharf 1997).

Simulation data in two cosmological scenarios have been used: a standard 
CDM (SCDM) model and a CDM + $\Lambda$ ($\Lambda$CDM) model, respectively. 
The SCDM simulation was discussed by KCOR94 and is characterized by 
the following parameters: spectral index for the initial power spectrum 
of perturbations $n$ = 1, normalized Hubble constant 
$h \equiv $ H$_0/(100$\kmsmpc$) = 0.5$, total mass density $\Omega_M = 1$, 
baryonic fraction $\Omega_b = 0.06$, and normalization $\sigma_8 = 1.05$. 
On the other hand, the \lcdm simulation was discussed by CO94 and 
featured the following parameters: $n = 1,~ h = 0.6, \Omega_M = 0.45, 
~\Omega_\Lambda = 0.55~ (\Omega_0 =\Omega_M +\Omega_\Lambda = 1), 
~\Omega_b = 0.043$ and $\sigma_8 = 0.77$. 
In both calculations a cubic region of size 85 \hinv Mpc at the current epoch
was simulated inside a computational box with $270^3$ cells and 
$135^3$ dark matter (DM) particles, allowing a spatial resolution 
of 0.315\hinv Mpc. We refer to KCOR94 and CO94 for further 
details. The simulations were performed with the code described in
Ryu \etal (1993). In addition to DM, the evolution of gas was followed 
with a grid-based hydrodynamic code.

\subsection{Analysis Method} \label{num2.s} 

A crucial and lengthy part in the current analysis of the results is 
the extraction and interpretation of detailed 
information from the numerical data. Each simulation form thousands of clusters
and a very rich system of associated shock waves.
We outline our methods here.

\subsubsection{Cluster Identification and DM Related Properties}

The first step to take is the identification of GCs in the data.
This could be done either through the X-ray emissivity criterion detailed 
in KCOR94 or through the distribution of DM particles. The former is 
preferred when the derived quantities must be directly compared with 
observational tests. However, when studying dynamical properties of GCs 
such as accretion velocity onto them, it would be better to use DM
particles for identification, because it is mainly the gravitational 
contribution of this component that determines those properties. 
Therefore, for the identification of GCs we have adopted the DM-based
``spherical over-density'' method described in Lacey \& Cole (1994) 
with a slight modification to speed up the calculation process. 
In particular, each DM particle is placed inside a cell (of physical size 
[0.315\hinv Mpc]$^3$) of the 
full 270$^3$ grid according to its position. If the number of particles 
inside such a cell exceeds a given threshold $N_{threshold}$, 
then, for each particle 
location, we define a local number density $n_{\ell} = 3 (N+1)/(4\pi R_N^3)$, 
where $R_N$ is the distance to the $N$th nearest neighbor. We then 
take the highest density particle locations as the candidate centers 
of clusters. The ``candidate'' particle locations are then sorted by 
density and a sphere is grown around each one of them, with the radius 
being increased until the mean density decreases to a value 
$\delta ~ \bar{\rho}$ where $\delta$ is a selected parameter and $\bar{\rho}
= ~\Omega_M ~ \rho_{crit}$ is the mean background density of the universe.
The center of mass (CM) of the particles inside the sphere is calculated 
and taken as a new center and the overall process iterated until 
center corrections are smaller than $\epsilon/n_{\ell}^{1/3}$, 
where $\epsilon$ is a small parameter. 
Once the cluster radii, $R_\delta$, are thus
defined, we reject from our list the smaller of any
two clusters whose spatial separation is 
shorter than ${3}/{4}$ the sum of their radii. 
For our analysis we choose $N_{threshold}=50$; as in 
Lacey \& Cole (1994) we set $N=10$ and $\epsilon = 0.1$. 
We chose $\delta = 80$, since then $\delta\rho$ is about the average density
inside the first caustic around clusters in SCDM (Bertschinger 1985; RK97).
So the cluster radius is defined by $R_{80}$. 

We point out for comparison that we have also identified GCs with the X-ray luminosity 
method outlined in KCOR94. In general we have found
the relations between various physical quantities are not affected 
significantly by the method used. However, since the criteria are different,
the samples considered are not identical, with differences 
depending on the thresholds used. 

\subsubsection{Cluster Baryonic Matter Related Quantities}

Once the cluster centers and radii, $R_{80}$, 
have been defined we turn to determine their
X-ray based temperature, $T_x$, and luminosity $L_x$; in addition we compute
the shock radius and velocity, $R_s$ and $v_s$ respectively, around clusters.
The cluster temperature $T_x$ is computed by averaging over 
the cluster core region defined by $r<R_{avg}=$ 0.5 \hinv Mpc.
The X-ray luminosity $L_x$, however, is averaged over a volume of $r<R_{avg}=$
1.0 \hinv Mpc, since bremsstrahlung emissivity $j_{ff}(\nu)$ (KCOR94, Eq. 2)  
is substantial up to a distance of $\sim1$ \hinv Mpc.
Operationally, the average is contributed from
all the the cells falling inside a sphere of volume $V$ (around the cluster), 
with a weight function, $w$, given by the cell intersection with $V$. 
So, for example, $w=1$ for central cells and $w\ll 1 $
for the marginal ones. This detail is important, because of the rapid drop of
the thermodynamic variables with distance from the cluster center.
Altering the ``natural'' weight of the central cells, unless motivated by a
corrective purpose, introduces a source of error in the calculated quantities.
While our temperature metrics should be fairly accurate, we
expect, on the basis of resolution studies (Cen \& Ostriker 1999b),
that the X-ray luminosities are systematically underestimated
while the rank order of the luminosities will be correct.

Once the cluster luminosity has been determined, we further select 
our sample based upon the criterion $L_x \ge L_{ff} = 
10^{41}~ \mbox{erg s}^{-1}$.
We note that given the different numerical schemes for the evolution of 
the baryonic and DM related variables, the position of the X-ray
luminosity peak and the DM-based CM of the same cluster 
may not correspond exactly. The shift is limited to one grid cell for the high
luminosity clusters, but can amount a few cells for the faint ones. 
Our combined choice of $N_{threshold}$ and of a limiting luminosity assures 
that the luminosity peak and 
DM-based CM do not differ more than 1 grid cell.

Finally, in order to determine the shock radius, $R_s$, and
velocity, $v_s$, we have labeled those cells where shocks are located. 
Shocks have been identified as compression regions ($\nabla\cdot {\bf v} < 0$)
meeting the following requirements for pressure ($P$) and velocity ($v$) 
jumps (index 1 and 2 refer to pre-shock and post-shock quantities):
\begin{eqnarray}
\frac{P_2 - P_1}{P_1} & > & 1.5  \label{pjump}  \\
v_1 - v_2 = 2~ \frac{c_1}{\gamma +1} \frac{M_1^2 - 1}{M_1}  & > & 0.87 ~c_1,
\label{vjump}
\end{eqnarray}
where $c_1$ is the pre-shock sound speed.
These criteria select shocks with Mach number $M_1 > \sqrt{3}$. 
However, once shocks have been identified with this method,
we adopt the temperature jump (across the shock) instead of the pressure in
order to calculate their properties. In fact,
when a shock is only a few computational cells apart from a 
cluster center, compression in excess to that produced by the shock wave can
result from the gravitational potential of the cluster itself. Given finite
numerical resolution, it becomes then impossible to separate the two 
effects (due to shock and gravitational compression). However, since the
cluster core is thought to be rather isothermal (Evrard \etal 1996), 
adiabatic gravitational
compression affects density, pressure and velocity substantially more 
than the temperature. Therefore, for cosmological simulations, 
shock properties based on the temperature jump are the most reliable. 
Our tests clearly support this point.

The condition enforced by Eq. \ref{vjump} does not add any physical property to
the selected shocks, with respect to Eq. \ref{pjump}. However, it is of great
advantage in our numerical effort to rule out compression regions simply 
due to the hydrostatic equilibrium inside the
cluster gravitational potential, and otherwise detectable by
Eq. \ref{pjump} as shocked cells. This scheme has proven successful in tests 
with shocks of known hydrodynamic properties for several different
applications. 
Once the shocks have been properly identified, the shock radius, $R_s$, 
is calculated by taking the average of the 6 distances to the
first ``shocked'' cells encountered along the coordinate directions 
($x,~y$ and $z$) from the cluster center.
In addition, the normal component of the pre-shock velocity is calculated
using the hydrodynamic shock-jump conditions (\eg Landau \& Lifshitz 1997).
We note that the normal component of the shock velocity
$v_{n1}$ and not $v_1$ is really the quantity of interest here.
Therefore in the following this is the quantity we refer to as the shock
velocity $v_s$.
In fact it determines the shock Mach number and flux of
kinetic energy available for particle
acceleration, the parameters most relevant
for production of CR populations.

\section{Results} \label{res.s}

\subsection{Morphological Structure of Cosmic Shock Waves}\label{morp.s}

Figs. \ref{sjdv.f} and \ref{ljdv.f} illustrate typical structures found
in the simulations. They represent slices of 0.315\hinv Mpc thickness
showing bremsstrahlung emission (gray-scale) with superposed contours of 
the shock compression factor ({\bf$\nabla \cdot v$}). 
In both cases shocks confine filamentary
structures where the temperature can vary from $10^4$ K to $10^7$ K. 
The most interesting feature, however,
is the complexity of shocks around GCs. 
In both cosmological models, in fact, the accretion flow develops {\it multiple
shocks} extending over a region of $5-10$ \hinv Mpc size around the GCs.
This is a generic feature of accretion flows there
and it is directly related with
the hierarchical process for cluster formation. In fact the merging of two
sub-structures into a single unit produces at least two main effects related to
shocks. {\it First}, shock waves of low Mach number are generated in the collision 
of the clusters' ICMs, which are commonly referred as merger shocks.
But {\it also}, part of the accretion shocks previously 
associated  with each sub-structure end up propagating through 
the ICM of the newly formed structure, reaching deep inside 
the cluster core. These shocks, which we refer to as {\it flow} shocks, are 
subsequently ``fed'' by residual gas motions in the ICM and ongoing
gas inflow accreting along filaments and sheets. 
Their presence, in addition to outer accretion shocks,
provides additional heating of the ICM and makes its thermal structure
not quite uniform over a region of several \hinv Mpc of size. 
As already pointed out in \S 1,
Markevitch \etal (1998 and also references therein) 
and Donnelly \etal (1999 and also references therein) showed evidences for
significant temperature structures inside clusters. 
They attribute them to the presence of shocks associated with merger events. 
However, in some cases the merging is
only inferred from the temperature map; also, in other cases temperature 
asymmetry is observed when the merger is just beginning to take place and
therefore has not been able to affect the cluster temperature structure yet. 
In the future cluster shocks could be identified 
independently of the presence of merging processes. Some of this evidence may
actually already be available 
(Ensslin \etal 1998, see \S 4 for further discussion).

A three dimensional perspective of shocks around GCs in SCDM
is offered in Fig. \ref{3ds.f}.
This is a combination of shock-strength isosurfaces (a)
and volume rendering of bremsstrahlung emission (b), 
for a portion of the computational domain of size 
$30 \times 40 \times 30$ (\hinv Mpc)$^3$ at $z=0$. With the help of 
the bremsstrahlung emission, which identify the GCs, one is 
allowed to locate shocks in their appropriate cosmological
context. This image reveals 
a further degree of complexity of real cosmic shocks with respect 
to the two-dimensional slices above.
Namely, in addition to being multiple associations with individual
clusters, such shocks are also largely connected topologically 
with neighboring structures. Their shapes, far from spherical, 
extend over tens of Mpc forming a continuum that
envelops all nearby clusters.

In conclusion, the hierarchical process for structure formation 
produces an extremely complex shock structure around clusters and
groups of clusters (or superclusters). 
{\it These shock waves are neither spherical
nor identifiable by a simple surface. Indeed they
intersect each other, forming nested shock surfaces, penetrating deep
inside the ICM of individual clusters.}

\subsection{Physical Relations for Accretion Shock Waves}
 \label{rel.p}

In this section we aim for a more quantitative description of accretion 
flows and shocks.
Fig. \ref{mh.f} includes four histograms showing 
the distribution, at redshift zero, of GCs (N[GC])
with respect to the Mach number ($M$) of the associated 
``external'' accretion shocks (top) and ``internal'' shocks (bottom). 
The latter have been selected inside regions of radius 0.5 \hinv Mpc 
from the cluster centers.
The results are plotted
on the left (right) for the SCDM ($\Lambda$CDM) model. 
Mach numbers associated with external accretion shocks are remarkably large,
ranging from $\sim 10$ up to a few $\times 10^3$ in both cosmologies.
We note that this is partly because photo-heating of pre-shock gas
by metagalactic radiation field was not included. In addition, 
feedback processes from massive stars in galaxies may raise the 
temperature in the outskirts of clusters further.
So without these sources of heating the pre-shock gas stays cool with
sound speed less than $\sim 1\kms$. If photo-heating was
properly included, however, the pre-shock gas temperature 
would have been raised to circa $\sim10^4$ K,
with corresponding sound speed $\sim 30\kms$ (see \eg Ostriker \& Cen 1996;
Cen \& Ostriker 1999a). 
Hence, the Mach numbers should
be significantly smaller than those in Fig. \ref{mh.f}. Yet they would still
reach up to $\sim 100$ for external accretion shocks.
On the other hand, the Mach numbers associated with internal shocks are much
smaller, mostly in the range $\sim 3-10$ and peaking
about 5. It thus emerges that, because internal shocks propagate
through a significantly hotter medium with typically $T\sim 10^6- 10^7$ K,
the velocities associated with
both external and internal shocks are comparable.

Among the characteristics of accretion flows onto GCs the most 
relevant quantities are the shock velocity, $v_s$, and radius, $R_s$. 
These are plotted as a function of each 
cluster's X-ray temperature in Fig. \ref{vrt.f}, again on the left
(right) for SCDM ($\Lambda$CDM). 
According to the self-similar solution of one-dimensional spherical
accretion in the SCDM universe, the cluster is confined by an accretion
shock at $R_s$ and the temperature of the cluster gas increases inward
(Bertschinger 1985; RK97).
By choosing the gas temperature at $r=0.3R_s$
as a representative value for the cluster's temperature $T_x$,  
the following relations of $v_s$ versus $T_x$ and $R_s$ 
versus $T_x$ are expected: 
\begin{eqnarray} \label{bert1.e}
v_s = ~1.75 \times 10^3~ \kms
\left(\frac{T_x}{7.8 \times 10^7~\mbox{K}}\right)^\frac{1}{2} 
\\ \label{bert2.e}
R_s = ~2.12~ h^{-1}~ \mbox{Mpc} 
\left(\frac{T_x}{7.8 \times 10^7~\mbox{K}}\right)^\frac{1}{2}.
\end{eqnarray}
By fitting our data for $v_s$ and $R_s$
to a function of the form 
\begin{equation}
f(T_x) = K ~\left(\frac{T_x}{7.8 \times 10^7~\mbox{K}}\right)^\alpha,
\end{equation}
as suggested by Eqs. \ref{bert1.e} and \ref{bert2.e}, we have 
obtained the values for the coefficients 
$K_{v_s},~\alpha_{v_s}$ and  $K_{R_s},~\alpha_{R_s}$ respectively, 
that are reported in Table \ref{par.t}. Note that 
$\alpha_{v_s}=\alpha_{R_s}=0.5$ is expected for {\it both} SCDM and 
$\Lambda$CDM from {\it scaling relations}.
Clearly the expected trends are reproduced only for the shock velocity,
whereas strong deviations from Eq. \ref{bert2.e} appear 
in the plots for \ra in both cosmologies (Fig. \ref{vrt.f} bottom panels).
The most dramatic are the differences for $\alpha_{R_s}$. The small value
reported in Table \ref{par.t} with respect to that in Eq. \ref{bert2.e}
suggests \ra is almost independent of the type of cluster. 
Although numerical errors both in the simulations and in the data
analysis must be considered, such discrepancies are probably true and
related to the actual structure of the flows. As already described
in the previous section, the formation process of GCs imprints complex,
{\it irreducible} three-dimensional shock structures which make it difficult 
to describe each GC with a single shock radius.

On the other hand, as long as the thermalization of the accretion 
kinetic energy takes place, the \va versus \tx relation
is less affected by the flow structure, with $\alpha_{v_s}$ very close to 0.5
for both cosmologies. $K_{v_s}$, is consistent with the
predictions of self-similar solutions for SCDM as well as for $\Lambda$CDM
due to the same reason.
We point out that $v_s$ is the normal component of the accreting gas in the 
shock rest frame. This is the component that undergoes dissipation 
originating the postshock gas temperature. However, the three-dimensional
accretion flow also possesses transverse velocity components
that are not thermalized across the shock and that can generate 
turbulent motions inside the IGM of GCs. In this regard, Eulerian, 
uniform grid-based schemes may be among the best choices to capture this
component of the flow in term of a balance between resolution and computational
performance. In fact, on the one hand in 
Smoothed Particle Hydrodynamic methods turbulence can be suppressed by
excessive viscosity. On the other hand, the higher computational cost paid by
the advantage of having a higher resolution with Adaptive Mesh Refinement
techniques is not completely satisfactory because 
the small scales of the turbulent component are not fully generated.

We have also calculated the flux of kinetic energy across
shocks defined as 
\begin{equation} \label{fl.eq}
\Phi_{E_k} = \frac{1}{2} ~\rho ~v_s^2 ~R_s^2 ~v_s .
\end{equation} 
This quantity is shown in 
Fig. \ref{ft.f} as a function of \tx for the SCDM (left) and \lcdm (right)
models. 
It is  of interest
in relation to CR acceleration, because it represents the amount of
power available for conversion into supra-thermal particles.
It provides a large amount of power 
of the same order of the X-ray luminosity emitted from a central region of
0.5 \hinv Mpc (see KCOR94 and CO94
for the amount of the cluster's X-ray luminosity).
It is clear that if a modest
fraction of this inflowing kinetic energy can be converted into CRs,
these may become important sources of emissions (by synchrotron and
inverse-Compton scattering) and even 
play a dynamical role through CR pressure.
$\Phi_{E_k}$ was fitted by a power law of the form
\begin{equation}
\Phi_{E_k} = K_\Phi 
\left(\frac{T_x}{7 \times 10^7~\mbox{K}}\right)^{\alpha_\Phi}.
\end{equation}
The best fit parameters are reported in Table \ref{par.t}. As we can see,
the normalization factor, $K_\Phi $, is larger for the \lcdm case than for the
SCDM one. Moreover, both slopes are larger than the values
implied from the slopes of \va and \ra in combination with Eq.
\ref{fl.eq}. 
This is probably due to an additional dependence of the accreting gas density
on the cluster temperature. In fact,
$\Phi_{E_k}$ is calculated at the first shock cells encountered 
along the coordinate axis from the GC center (and then averaged over the
accretion surface). Since such cells could well be inside ``external''
shocks, as illustrated in Figs. \ref{sjdv.f} and \ref{ljdv.f},
the density of the accreting gas depends on
the properties of the cluster environment. In particular we know that: 
(1) for a given cosmological model 
such density is higher around larger (higher temperature) clusters, implying a
steeper increase of $\Phi_{E_k}$ than for a temperature independent density
value. Also, (2) for
the same cluster temperature, the corresponding cluster gas density and,
therefore, $K_\Phi$ are larger for a \lcdm model than in a SCDM one.
This second point is related to the fact that in general clusters in \lcdm
have temperature smaller that those in SCDM (\eg Figs. 5 and 6; see also
KCOR94, CO94).

It turns out that $\Phi_{E_k}$ is a steep function of
cluster temperature, spanning several orders of magnitude in 
the temperature range of the identified clusters. 
This means that if CR acceleration mechanism at
shocks around GCs possesses an injection mechanism and
an efficiency independent of the cluster properties (\eg mass and 
temperature), then {\it we would expect hotter clusters to store a 
relatively larger amount of nonthermal energy in the form of relativistic 
particles}. Such trend has been observed already: for instance, Liang 
(1999) reported a positive correlation between the radio emission and 
the X-ray temperature in GCs.

\subsection{Evolutionary Trends} \label{nmz.s}

In general, complex shock structures are already present at high redshift.
At $z=5-10$ shock waves are well formed and have already developed 
connections with neighbor clusters or protoclusters.
The strength of the shocks is largest around the most massive
objects, yet far from uniform. As the evolution advances, mergers occur 
on all scales, affecting all types of structures including shocks. 
As a result, at $z=0$, many filamentary structures
have coalesced into larger ones. 
In addition the shocks associated with them have
become stronger and more uniform, due to the increased amount of matter
onto which the gas is being accreted.

In order to study the characteristics of shock evolution
we define the following quantity:
\begin{equation}
S(z,M) = \frac{1}{N_{tot}(1+z) dx}
\int^M_\infty \frac{d}{dM^\prime} (N_{shock}[z,>M])~ dM^\prime
\end{equation}
Here $N_{shock}(z,M)$ is the number of cells hosting a shock of Mach number
greater than $M$ at redshift $z$, $N_{tot}$ the total number of cells in the 
computational box and $(1+z)dx$ is the comoving cell size.
Fig. \ref{nmz.f} shows the evolution of $S(z,1)$,
representing the inverse of the average
comoving distance between shocks of any Mach number,
as a function of redshift ($z$) for SCDM
(open circles) and \lcdm (solid circles). This
quantity also represents the ratio of shock surface over space volume.
The following characteristics stand out from this plot. First, 
it is clear that cosmic plasma was
populated with many more shocks in the past than nowadays. 
Second, while $S(z,1)$ peaks at $z \simeq  4.6$ in the
$\Lambda$CDM model, we can only say that it peaks at $z \gtrsim  5$ 
in the SCDM model, given our limited data-set. 
Also, in recent epochs $S(z,1)$ is substantially
larger in the $\Lambda$CDM model than in the SCDM model.
Finally, the evolution is smooth in  the $\Lambda$CDM case,
but shows abrupt transitions in the SCDM one.

As for the first point made above,  
the larger area of shock surfaces
is due to the extremely filamentary structure of the universe 
at higher redshifts. Filaments are confined 
by accretion shocks. As already pointed out in \S \ref{morp.s}, 
as structure formation progresses, filaments coalesce,
therefore growing thicker and rarer. 
Although the size of their associated shocks increases, 
their reduced population plays
the dominant role determining overall a decrease of the total 
area of shock surfaces in a comoving volume.

The second point then tells us that
although such a process takes place in both cosmologies, there are 
many more filaments today in the \lcdm than in the SCDM scenario.
This is clearly illustrated in Fig. \ref{tmp.f}, 
where we show two-dimensional slices, with thickness 0.315\hinv Mpc, of
the temperature structure of the universe at two different redshifts, $z=3$
(a) and $z=0$ (b), for SCDM (left) and \lcdm (right) models. There
we can see that at $z=3$ the two universes look quite similar, with comparable
amounts of filaments, in accord with Fig. \ref{nmz.f}. However, at later
times, \eg $z=0$, filaments in the SCDM model are fatter and rarer as
compared to the \lcdm case.
This finding is a reflection of the different initial conditions in the two
models. In particular, the larger amplitude of the primordial perturbations
used in the SCDM, $\sigma_8=1.05$, gives rise, at current epoch, to a more
clustered but less filamentary structure in this model than in \lcdm
(\cf KCOR94; CO94).

Turning to the final point, sudden reductions
of $S(z,1)$ are located
at $z\simeq 3,~0.7$ and 0.2 in SCDM.
These reflect the occurrences of a higher rate of merging processes at
those particular epochs. 

Further details related to the time evolution of shocks are
illustrated in Fig. \ref{wf.f}, where we plot as a function of 
the Mach number the quantity
\begin{equation}
W(z,M) = \frac{d}{dz} S(z,M).
\end{equation}
$W(z,M)$ expresses the negative of
the rate of formation of shocks with Mach number 
greater than $M$ at a particular epoch $z$. 
Thus, from Fig. \ref{wf.f} we can see that at early times ($z=4.5$)
shocks are forming with Mach number in the range between circa 
10 and a few $\times 10^2$ for both \lcdm and SCDM cases.
No shocks exist with $M>$ a few $\times 10^2$ and for 
$M<10$ shocks are being depleted due to merger events.
Later on, at $z=1.25$, shocks start forming in the range
$10^2 \lesssim M \lesssim 10^3$. 
In accord with Fig. \ref{nmz.f}
the total rate of shock formation ($\equiv W(z,1)$) is always negative.
As already pointed out, the numerous weak
filament shocks are replaced by stronger but rarer shocks.
The last two panels ($z=0.6$ and $z=0.05$ respectively)
show the smooth shock formation evolution in the \lcdm.
A small amount of shock formation still occurs for high Mach numbers
but overall the shock population is decreasing at an increasing rate.
The situation is more complex in the SCDM case.
At $z=0.6$, identified above as an epoch of high merging rate,
shocks are reduced at any Mach number. 
At $z=0.05$ the largest Mach number shocks ($M \gtrsim 5\times
10^2$) are depleted indicating merging of the most massive objects.

In order to assess the relative importance of shocks at different epochs
in terms of CR contribution, in Fig. \ref{phst.f} we plot the adimensional 
quantity
\begin{equation}
F(z) = \frac{(\Phi_{E_k}[z])_{shock}}{(\rho_c H_0^3 \lambda_{NL}^5)_{z=0}} 
\end{equation}
where $\Phi_{E_k}$ is the total flux of
kinetic energy through shocks (of any Mach number), $\rho_c$ is the critical 
density, $H_0^3$ the Hubble constant and 
$\lambda_{NL}$ the non-linear perturbation wavelength set to 50 \hinv Mpc.
Fig. \ref{phst.f} shows that the flux of kinetic energy through shocks
today has increased by a few order of magnitudes with respect to early epochs,
say $z\simeq 5$. Thus, today's shocks retain more kinetic energy than ever.
However, the time integrated flux of kinetic energy 
through shocks (\eg since $z=5$) is much larger than the thermal energy 
content at $z=0$, which, according to our data, has mostly been
produced after $z\sim 1.5$.
This is not due to low thermalization efficiency of shocks. Rather,
that is because, although shocks form at much
higher redshifts, the thermal energy they produce undergoes severe adiabatic
losses due to cosmological expansion.
On the other hand, after roughly $z\sim 1.5$, such thermal energy is
retained inside well formed structures such as clusters and filaments.

In order to identify the characteristics of the most relevant
shocks, \ie those that process most of the gas, we calculate  
\begin{equation}
Y(M) = \frac{1}{E_{th}(z=0)}\int_{t(z=1.5)}^{t(z=0)} \frac{d\Phi_{E_k}(M)}{dLogM}
dt^\prime ,
\end{equation}
where the extremes of integration have been chosen on the basis of
the arguments in the previous paragraph.
Here, $\Phi_{E_k}(M)$ is the kinetic energy flux through 
shocks with Mach number between Log$(M)$ and Log$(M+$d$M)$, and 
$E_{th}(z=0)$ is the total thermal
energy inside the computational box at $z=0$.
This quantity is plotted in Fig. \ref{jpo.f}, as a function of Mach number $M$,
in the right (left) panel for the SCDM (\lcdm) model.

We can see that most of the flux of kinetic energy occurs through shocks
with (``low'') Mach numbers around 4 which correspond to our internal shocks. 
In fact, although most of the shocks have
Mach numbers much larger than that (see histograms
of Fig. \ref{mh.f}), low Mach number shocks are typically located inside
much denser regions (formed structure) and therefore process much more
matter and kinetic energy than on average
(as already pointed out in \S \ref{rel.p} the gas velocities for
internal and external shocks are comparable).
This depiction is not just the result of the integration.
A more detailed analysis of plots of $dY(M)/dt$,
describing the flux of kinetic
energy as a function of $M$, 
for different redshifts (not shown here) shows that the
the qualitative features of the curve in Fig. \ref{jpo.f}
are common at any $z$ 
and therefore the low Mach number shocks are always responsible
for most of the processing of the kinetic energy of the peculiar motions.

Finally, integration of the area underneath each curve in Fig. \ref{jpo.f} 
represents the total kinetic energy 
passed through shocks since $z=1.5$ divided by the thermal energy at $z=0$.
Its value is $\sim 17$ for the SCDM and $\sim 13$ for the \lcdm model
respectively. 
If a fraction $\epsilon\sim 10^{-2}$ of such energy is transferred to CR
protons, then 
the energy stored up in CRs today should amounts to
about 15 \% of the thermal energy inside formed
structure. This is only a rough estimate which needs to be refined by more
accurate calculations.

\section{Discussion} \label{dc.s}

We have studied the properties of ``cosmic shock waves'' associated with 
the large scale structure of the universe in two different cosmological
scenarios, namely SCDM and $\Lambda$CDM. 
Such shocks reveal remarkable properties. 
In fact, hierarchical formation histories of GCs produce highly
complex flows and shock structures, which extend over scales of several Mpc.
In addition to accretion shocks (responsible for heating 
infalling gas) and merger shocks, {\it flow} shocks also appear 
and propagate through 
the thermalized ICM, providing extra gas heating. 
It turns out that the morphology of shocks associated with 
a large scale structure is complex and 
irreducibly three-dimensional and 
spherical shapes are inadequate to their description.
Only  for the external accretion shocks, located far away from the cluster core, 
some form regularity is recovered. This is an important issue especially in
perspective of those missions with the next generation
of high resolution X-ray telescopes (Chandra and XMM)
which are planning to detect shocks in the ICM.
It is worth mentioning 
that Ensslin \etal work (Ensslin \etal 1998) might already provide
observational evidence for the presence of flow 
shocks in cosmic structure. Their conclusions are based on the assumption 
that the observed radio emission is due to particles currently 
accelerated at shocks there and injected from a ``radio relic'', 
a remnant previously associated with some radio galaxies. 
For example, for 1253+275, 
they find a pre-shock gas temperature $T\sim 0.5-1$ KeV which shows that
this is not the case of an accretion shock but that of 
either a merger or a flow shock
(propagating through the ICM). 
Since there is no evidence for a merging process
in 1253+275, it must be a flow shock. In addition, from their 
reported data on the pressure jump we infer $M\sim 3 - 4.2$, well 
in the expected range for the internal shocks shown in Fig. \ref{mh.f} 
(bottom panel).

Cosmic shocks are also ideal sites for particle acceleration.
We have shown already in \S \ref{rel.p} that cosmic shocks provide enough 
power to produce copious CRs. The details of the produced populations 
will depend on the injection mechanism and scattering agent, \ie the 
magnetic field and the diffusion properties. Here, we further stress the
important role of merger and flow shocks. 
These shocks may be responsible, not only for
acceleration of CRs out of the thermal pool of the ICM, but also for the
re-acceleration of CRs produced at accretion
shocks and/or ejecta from radio-galaxies, AGNs or normal galaxies.
In addition, they could be crucial in terms of acceleration and
transport of ultra high energy CRs, because the scattering mean free
path of these particles is of the same order as the typical 
separation between accretion/merger/flow shocks in the ICM.
As already pointed out at the beginning of this section, there seems 
to be solid foundation for the existence of such shocks.

As described in \S 1, the presence of
relativistic CR electrons in GCs has been inferred through
observations of diffuse synchrotron radiation from radio halos.
In addition, there is evidence for 
excess of radiation in both EUV and hard X-ray bands,
with respect to thermal emission.
Although still of controversial interpretation, such excesses are probably due
to IC emission of CR electrons scattering off cosmic microwave background
photons. Published studies, however, reveal that for an accurate interpretation
of the constraints from the combined non-thermal emission components, 
it is crucial to have a detailed depiction of the relative
distribution of particles and magnetic fields (\cite{elb99}).

The proton component of
CRs has not been directly observed. Nonetheless, given the estimates for the 
CR electron component, Lieu \etal (1999) concluded that 
their contribution in terms of 
dynamical pressure in GCs could be comparable to the thermal gas.
This is consistent with the estimate inferred in the previous section and, as
already pointed out in the introduction, has important consequences for
cosmology. 
We point out that cosmic shock waves have existed ever since 
nonlinear structure formation
was initiated at high redshift. This was shown through Fig. \ref{nmz.f}
in \S \ref{nmz.s}. Therefore, the importance of nonthermal activities in the 
cosmic plasma traces back to early epochs.
Collisions of CR proton in the ICM generate, however, a flux of gamma ray
photons through the production and subsequent decay of neutral pions;
as pointed out by Blasi (1999), such gamma ray flux seems to be only 
marginally compatible with the 
upper limits measured by EGRET for Coma and Virgo clusters. 
But, again, the spatial and spectral distribution of CR, both
depending on the overall cosmological 
history of these particles, play a crucial role in
the determination of the expected gamma ray flux. 
In any case,
the advent of the new generation of $\gamma$-ray facilities
(GLASS, VERITAS) characterized by a much 
higher sensitivity (\cf Blasi 1999 for more details)
will definitely settle the issue.

From this depiction it emerges the importance and the necessity to
understand the role of CRs in cosmology. 
For this purpose we are developing numerical tools in order to treat
consistently 
magnetic fields and CRs in numerical simulations of structure formation.
Such tools, in fact, will allow us to follow explicitly the evolution 
of the magnetic field as well as the acceleration and transport of CRs.
With this information we will be able to carry out very 
useful comparisons between numerical and observational
results in various bands of the electromagnetic spectrum.

\acknowledgments
FM was supported in part by a Doctoral Dissertation Fellowship 
at the Uiversity of Minnesota. FM and TWJ were supported in part by NSF 
grants AST9616964 and INT9511654, 
NASA grant NAGS-5055 and by the Minnesota Supercomputing Institute.
DR and HK were supported in part by grant 1999-2-113-001-5 from the
interdisciplinary Research Program of the KOSEF. RC and JPO were
supported in part by NSF grants AST-9803137 and ASC-9740300.

\clearpage
\begin{deluxetable}{ccccccc}
\footnotesize
\tablecolumns{7}
%\tablewidth{294pt}
\tablecaption{Accretion Flow Best Fit Parameters \label{par.t}}
\tablehead{  
\colhead{MODEL} &
\colhead{$K_{v_s}$} &
\colhead{$\alpha_{v_s}$} &
\colhead{$K_{R_s}$} &
\colhead{$\alpha_{R_s}$}  &
\colhead{$K_\Phi$} &
\colhead{$\alpha_\Phi$} \\
\colhead{} &
\colhead{(10$^3$~$\kms$)} &
\colhead{} &
\colhead{(Mpc)} &
\colhead{} &
\colhead{($10^{45}$ erg s$^{-1}$)} &
\colhead{}
}
\startdata
SCDM     & 1.8   & 0.47  &  1.75  &  0.1 & 2 & 2.0 \nl
\lcdm    & 1.9   & 0.52  &  1.4  &  0.1 & 6.7 & 2.0 
\enddata
\end{deluxetable}

 \clearpage

\begin{center}
{\bf FIGURE CAPTIONS}
\end{center}

\figcaption[]{
Typical cosmological structure. Bremsstrahlung X-ray emission (gray-scale) 
superposed to contours of the compression factor of shock waves 
(${\bf \nabla \cdot v}$) in SCDM is shown.
\label{sjdv.f}}

\figcaption[]{Same as Fig. \ref{sjdv.f} but in $\Lambda$CDM.
\label{ljdv.f}}

\figcaption[]{Three-dimensional 
shock surfaces (a) and
volume rendering of bremsstrahlung X-ray emission (b) 
for a SCDM model simulation. The colors have been assigned according to a
rainbow-type colormap where high values correspond to 
blue/violet and low ones to red.
The displayed frame includes a portion of the computational
box of size $30 \times 40 \times 30$ (\hinv Mpc)$^3$ at $z=0$.
\label{3ds.f}}

\figcaption[]{Histograms showing the distribution of clusters of galaxies
versus shock Mach number. For external, accretion shocks (top) and those inside
a region of 0.5\hinv Mpc radius around the cluster center (bottom) and for 
SCDM (left) and \lcdm (right).
\label{mh.f}}

\figcaption[]{Accretion velocity, $v_s$ (top), and radius, $R_s$ (bottom), 
as a function of cluster core temperature, $T_x$, for 
SCDM (left) and \lcdm (right).
\label{vrt.f}}

\figcaption[]{Kinetic energy flux across accretion shock, $\Phi_{E_k}$, as a
function of cluster core temperature, $T_x$. Again 
SCDM ($\Lambda$CDM) is on the right (left).
\label{ft.f}}

\figcaption[]{$S(z,1)$, representing
the inverse of the
average comiving distance between shocks,
as a function of cosmological redshift $z$ for SCDM (open circles) and
\lcdm (solid circles).
\label{nmz.f}}

\figcaption[]{Slices of temperature distribution showing the filamentary
structure in SCDM (left) and \lcdm (right) models at cosmological redshift
$z=3$ (a) and $z=0$ (b).
\label{tmp.f}}

\figcaption[]{Redshift derivative of the fraction of shocks with
Mach number greater than $M$ as a function of $M$. Note that, since 
$dt\propto -dz$, positive values of $W$ indicate a decreasing population
of shocks.
\label{wf.f}}

\figcaption[]{Evolution of $F(z)$ representing
the total kinetic energy flux through shocks of any Mach number,
normalized to $\rho_c H_0^3 \lambda_{NL}^5$. Here
we set $\lambda_{NL} = 50 $ \hinv Mpc. 
Open dots are for SCDM and filled dots for \lcdm case.
\label{phst.f}}

\figcaption[]{Time-integrated amount of kinetic energy passed thorough
shocks with Mach number between Log($M$) and Log($M+$d$M$),
since the epoch at $z=1.5$, normalized to the total thermal
energy inside the computational box at $z=0$ and divided by dLog$(M)$
(=0.1), as a function of Mach number $M$.
Plots are for SCDM (right) and \lcdm (left) models respectively.
\label{jpo.f}}

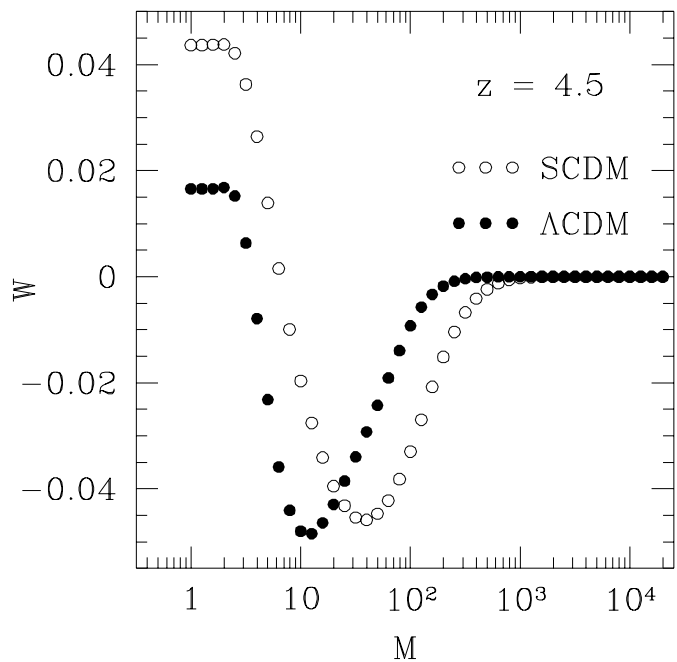
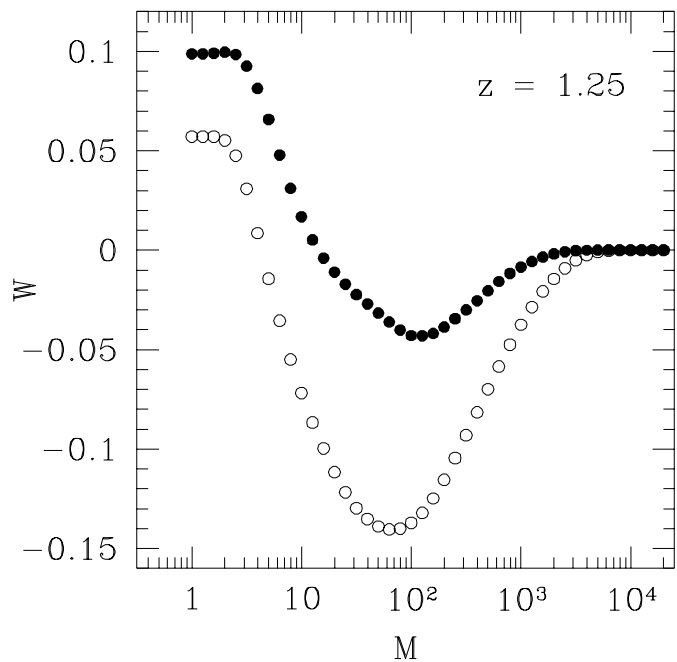
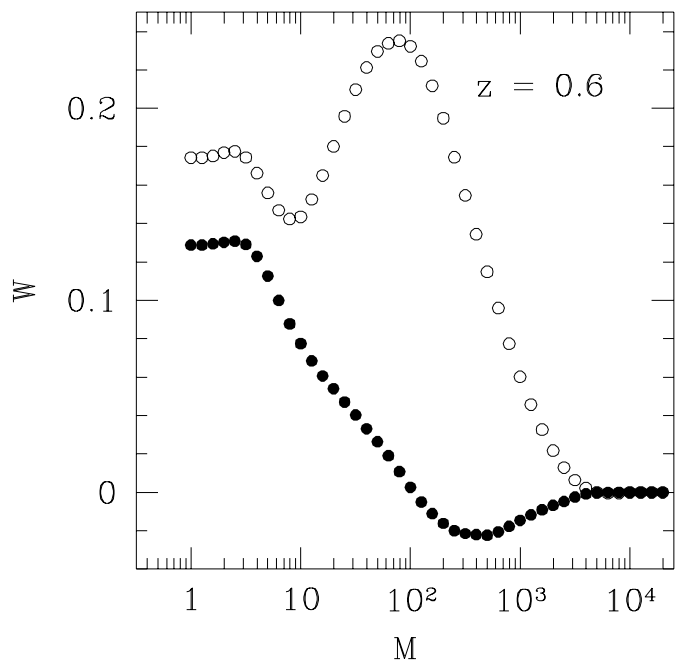
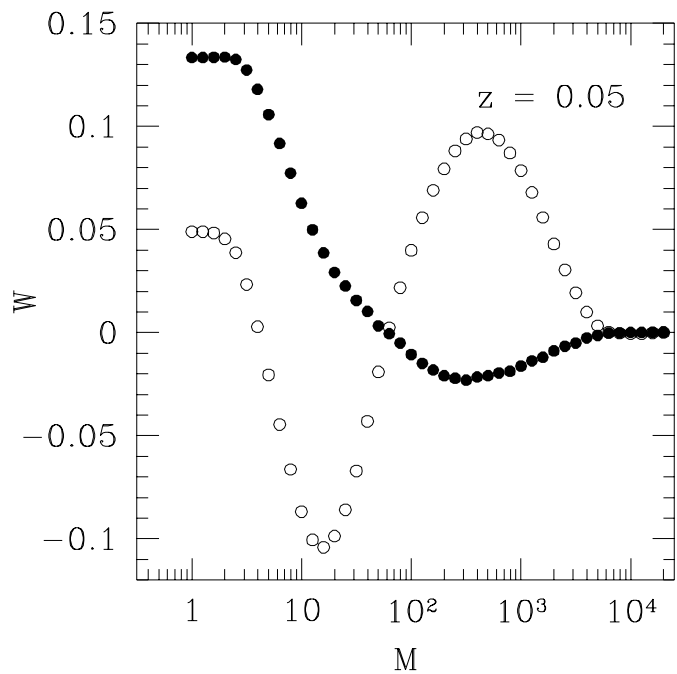
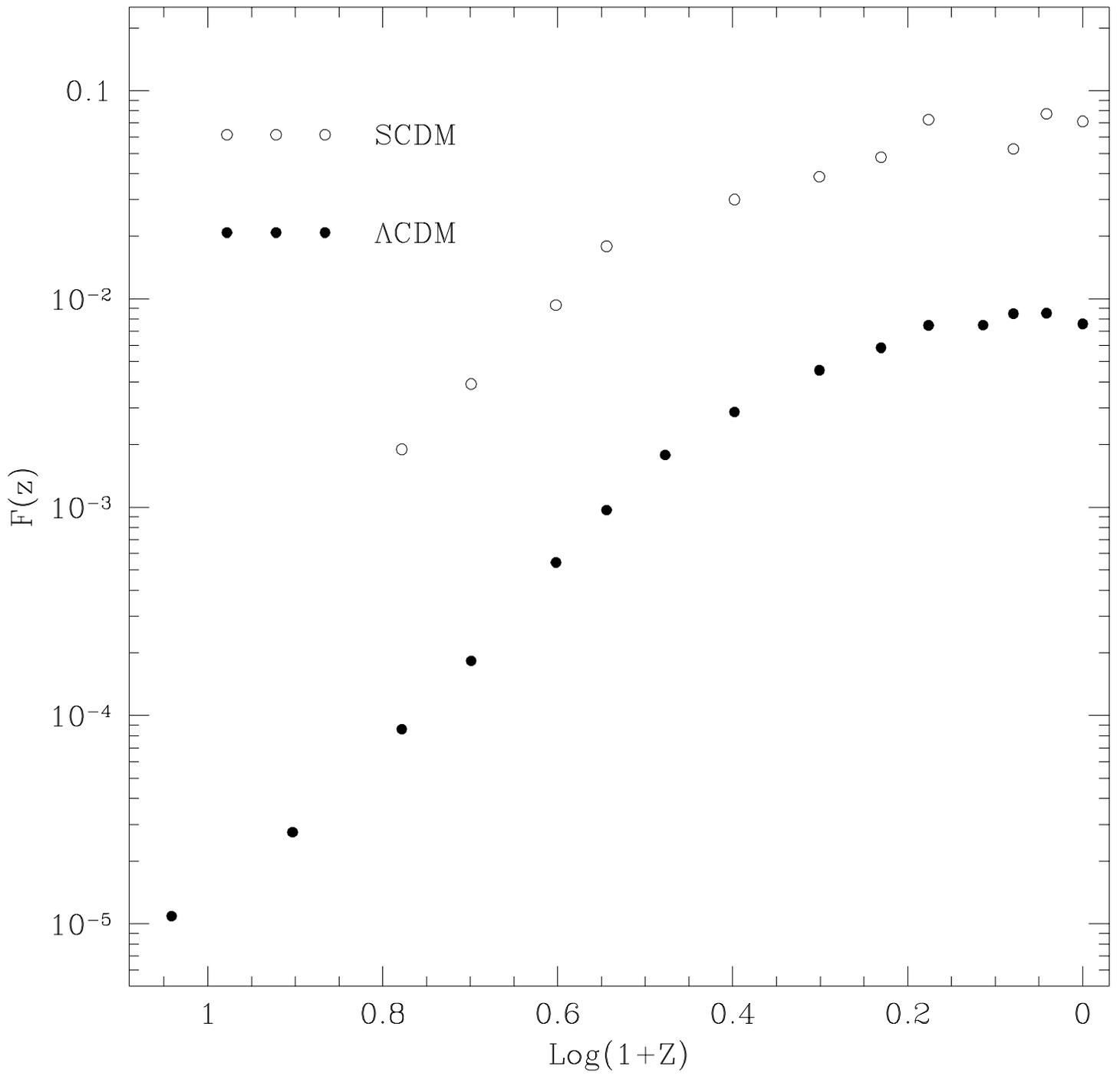
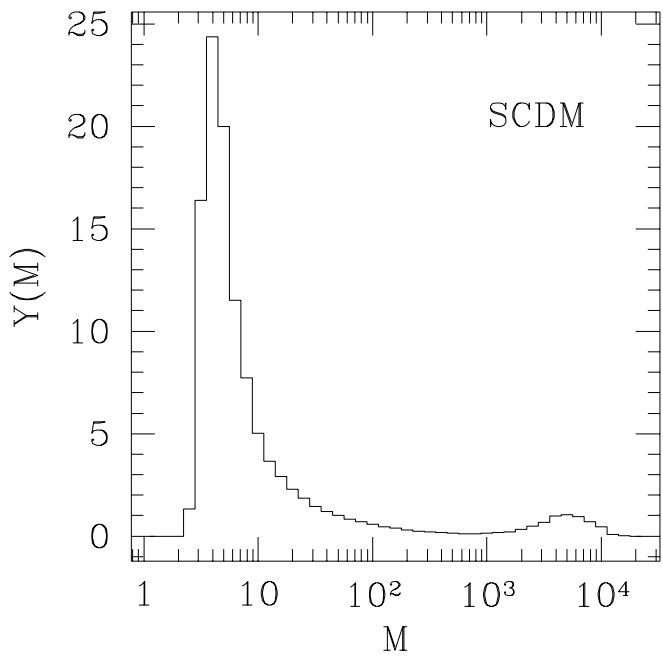
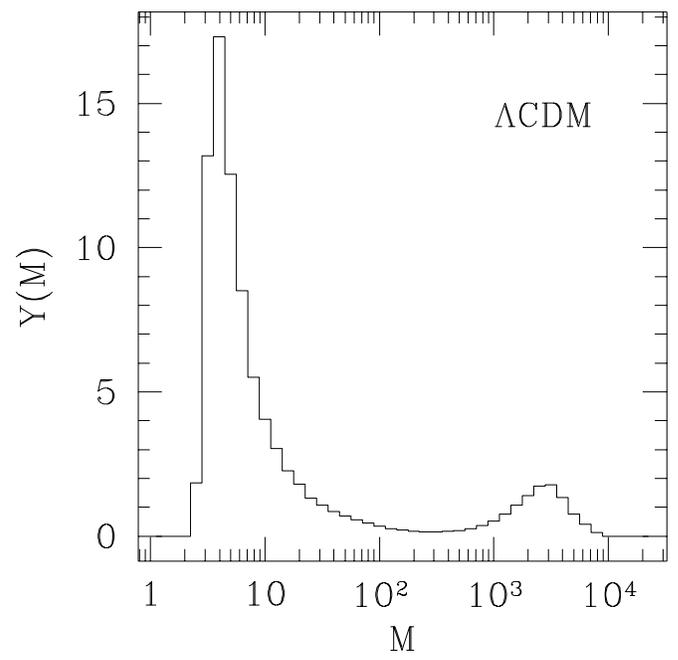
\end{document}